\documentclass[12pt, twocolumn,showpacs,preprintnumbers,aps,prl,reprint,superscriptaddress]{revtex4-1}

\usepackage{graphicx}
\usepackage{subfigure}
\usepackage{color}
\usepackage{amsmath, amsfonts, amssymb}
\usepackage[linktocpage,colorlinks=true,linkcolor=blue,citecolor=blue,breaklinks=true]{hyperref}
\usepackage{breakcites}
\usepackage{pgfplots}
\newcommand{\be}{\begin{equation}}
\newcommand{\ee}{\end{equation}}
\usepackage{natbib}
\usepackage{nameref}
\usepackage{xcolor}
\usepackage{cleveref}

\begin{document}

\preprint{}

\title{Coupling functions in climate}

\author{Woosok Moon}
\affiliation{Department of Mathematics, Stockholm University 106 91 Stockholm, Sweden}
\affiliation{Nordita, Royal Institute of Technology and Stockholm University, Stockholm 106 91, Sweden}

\author{John S. Wettlaufer}
\affiliation{Yale University, New Haven, CT 06520-8109, U.S.A.}
\affiliation{Nordita, Royal Institute of Technology and Stockholm University, Stockholm 106 91, Sweden}

\date{\today}

\begin{abstract}
{We examine how coupling functions in the theory of dynamical systems provide a quantitative window into climate dynamics.  
Previously we have shown that a one-dimensional periodic non-autonomous stochastic dynamical system 
can simulate the monthly statistics of surface air temperature data. Here we expand this approach to two-dimensional dynamical systems to include interactions between two sub-systems of the climate.  
The relevant coupling functions are constructed from the covariance of the data from the two sub-systems. We demonstrate the method on two tropical climate indices; 
The El-Ni\~{n}o--Southern Oscillation (ENSO) and the Indian Ocean Dipole (IOD), to interpret the mutual interactions between these two air-sea interaction phenomena
in the Pacific and Indian Oceans. The coupling function reveals that ENSO mainly controls the seasonal variability of the IOD during its mature phase. This 
demonstrates the plausibility of constructing a network model for the seasonal variability of climate systems based on such coupling functions.}
\end{abstract}


\maketitle

\section{Introduction}

Climate reflects of a myriad of interactions operating over a wide range of time-scales. 
The spatially inhomogeneous distribution of shortwave radiative flux drives the atmosphere and ocean fluid-dynamically, leading to long-ranged communication through fluid advection and wave propagation \cite{1}.  Thus, generally speaking, a fluctuation in forcing in one region can have a response in another region that is exhibited 
 in the spatio-temporal statistical correlations between the time-series of climate variables.   These correlations contain fundamental characteristics of climate subsystems by revealing the underlying mechanisms of cause and effect.  Therefore, a central problem in climate dynamics is to
understand how to construct a theory containing the key interactions that couple distant regions on the planet. 

Since the advent and ready availability of large-scale computation, the term ``model'' generally appears to be synonymous with Global Climate Models (GCMs), which operate by numerically solving the conservation of mass, momentum and energy throughout the atmosphere/ocean/ice system with parameterizations 
representing sub-grid scale physics \cite{2}. Contemporary approaches to improve GCMs use data assimilation \cite{3}
and test and implement high-resolution schemes \cite{4}.  Due to the complex nature of climate (particularly fluid) systems, GCMs are amongst the most sophisticated numerical models developed.  Nonetheless, despite the enormous progress made in the development of GCMs, completely resolving the spatiotemporal processes in the 
climate system is far from satisfactory \cite{5,6}. Furthermore, because of the complex structure and high dimensional data produced by GCMs, 
 {\color{black} it is not straightforward to extract the dominant physical processes on multiple time scales with the aim of interpreting their mutual interactions.}

A complimentary approach that naturally treats the large separation of process time scales arises by appeal to the analogy with the stochastic dynamics of
Brownian motion \cite{7}.  In this spirit, we introduced \cite{8,9} a periodic non-autonomous stochastic Langevin equation
to simulate the statistics of monthly averaged time-series of surface air temperature; $\dot{x}(t)=a(t)x(t)+N(t)\xi$, where $a(t)$ and $N(t)$ are periodic functions with annual periodicity, and $\xi$ white noise. The approach {\color{black}effectively} simulates the second-order statistical moments of observations. 
Of particular note is that the seasonal evolution of the variability of a climate variable can be naturally interpreted in terms of the interaction between the {\em seasonal stability}, $a(t)$, and the {\em noise forcing}, $N(t)\xi$, which enables us to isolate physical processes responsible for the stability, such as the sea ice albedo feedback \cite{10} for the time-evolution of sea ice thickness or
the Bjerknes feedback \cite{11} for ENSO \cite{12}.

When a subsystem is influenced by non-local processes on a characteristic time scale(s), theoretical extensions to higher dimensions are {\color{black}required}.  
A classical framework to treat spatio-temporal variations is that of teleconnections, although there is no means within it to deal with multiple time scales rigorously.  Hence, we consider multi-dimensional stochastic dynamical systems. The idea of coupling functions, which have been used to 
reveal the interactions underlying processes such as synchronization \cite{13}, appear as a natural framework to capture such non-local effects. 
Thus here we develop a two-dimensional stochastic model containing coupling functions that represent a mutual interaction of  two variables. We will evaluate the efficacy of the approach on seasonal time-scales.  In particular, we focus on two tropical {\color{black}indices}, the Nino3.4 index \cite{14} and the 
IOD index \cite{15} representing the El-Ni\~{n}o Southern Oscillation (ENSO) and Indian Ocean Dipole (IOD) \cite{16}, 
respectively.  We use these data to construct a two-dimensional periodic non-autonomous stochastic model that then simulates seasonal variability and interactions. 
Through quantitative and qualitative comparisons between the model and the data,  we evaluate of the model and provide physical interpretations of the interactions
in ENSO and IOD. This should {\color{black} be} a basic step toward a climate network model for seasonal prediction.

\section{Climate: Intrinsically Periodic \& Non-autonomous}

Since Fourier's studies of the Earth's energy budget, shortwave solar radiation has been known as the central input of energy into the climate system.  
Indeed, it acts as the largest annual {\em external} periodic thermal forcing to the Earth system.  
For this reason, despite mathematical difficulties, periodic non-autonomous equations represent the time-evolution of climate systems on seasonal time-scales. 

The minimal form of energy balance models pits the incoming shortwave and outgoing longwave radiative fluxes against each other as \cite{17},
\begin{align}
C_P\frac{\partial T}{\partial t} = S_0(1-\alpha)-\bar{\sigma} T^4,
\end{align}
where $T$ is the temperature of Earth surface, $C_P$ is its effective heat capacity, $S_0$ and $\alpha$ {\color{black}the shortwave radiative flux}
and surface albedo respectively, and $\bar{\sigma}$ the Stefan-Boltzmann constant. In equilibrium we have 
\begin{align}
\bar{\sigma} T^4_E = S_0(1-\alpha),
\end{align}  
which gives us the equilibrium temperature $T_E=\sqrt[4]{S_0(1-\alpha)/\bar{\sigma}}$. 
If we assume that the temperature $T$ is not far from $T_E$, so that $T=T_E+x$ with $|x| \ll |T_E|$, then the time-evolution of $x$ 
under the influence of white noise, $\xi$, with constant amplitude ${\sigma}$, is represented by an Ornstein-Uhlenbeck process,
\begin{align}
\frac{dx}{dt} = -\lambda x + {\sigma}\xi,
\end{align}
where 
\begin{align}
 \lambda = \frac{4\sigma T^3_E - S_0|\frac{\partial\alpha}{\partial T}|}{C_P}, 
\end{align}
where $\lambda$ represents the overall deterministic stability of the climate relative to the equilibrium temperature of $T_E$. 
The albedo sensitivity, $\partial\alpha/\partial T$, is negative, thereby exhibiting positive feedback, whereas 
the sensitivity of the outgoing longwave radiative flux is positive, thereby stabilizing deviations from the equilibrium temperature $T_E$. 
The high-frequency fluctuations, such as those associated with weather, are represented using white noise $\xi$ with amplitude $\sigma$.
Such a class of autonomous Ornstein-Uhlenbeck processes was introduced in climate dynamics by Hasselmann \cite{7}, and is a rationale for ubiquitous red noise spectra, particularly for sea surface temperature (SST). 

Under certain circumstances it may be reasonable to view some climate subsystems as autonomous, but most of the components are forced in some manner and thus are intrinsically non-autonomous.  For example, on long time scales {\color{black}the shortwave radiative flux} $S_0$ is a periodic function, and quantities such as incoming longwave radiation, sensible and latent heat fluxes and advection, will vary on a range of time scales.   Indeed, we know from common experience that most climate variables have seasonal cycles, but the time scale of weather events is much shorter.

\begin{figure}[!h]
\centering\includegraphics[width=3.3in]{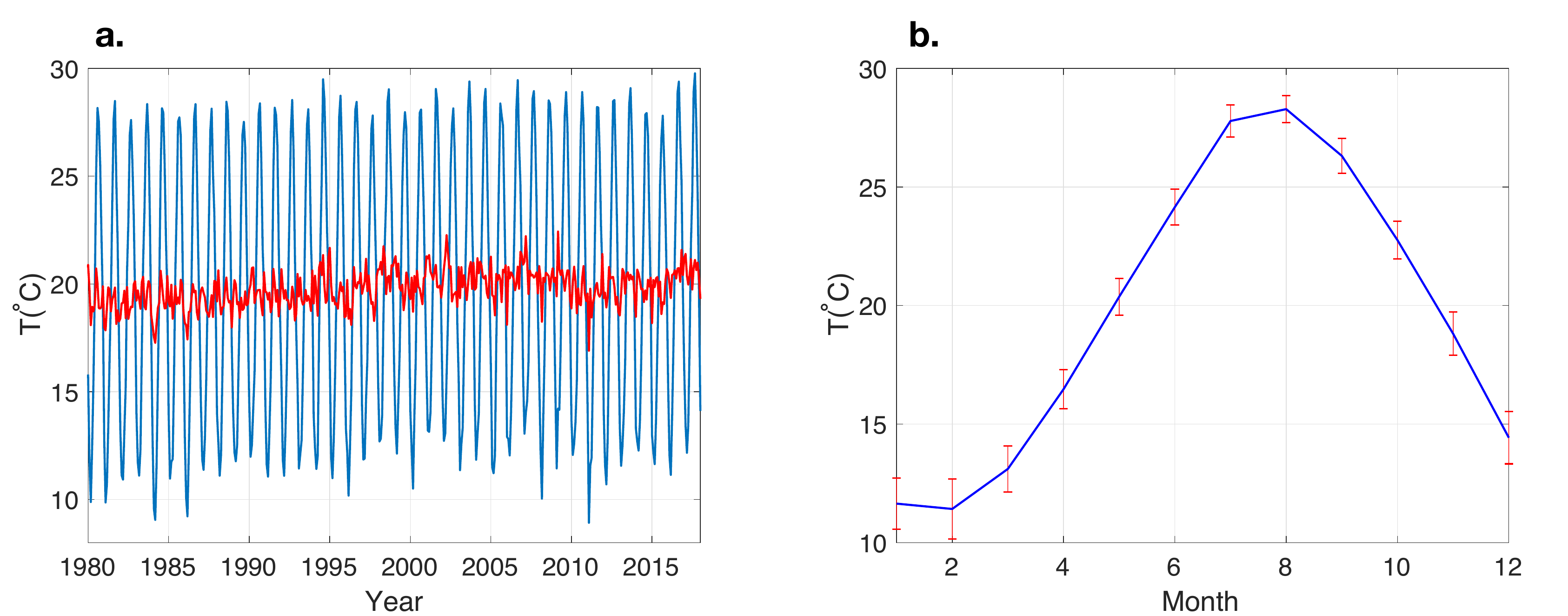}
\caption{A time-series of surface air temperature. (a) The original time-series (blue line) and the deviation from the climatological seasonal cycle 
plus an annual mean value (red line). (b) The climatological mean seasonal cycle (blue line) with monthly standard deviations
(red error bars).  }
\label{fig01}
\end{figure}

Figure \ref{fig01}(a) shows a time-series of surface air temperature (blue) from 1980 to 2018, exhibiting the dominant seasonal cycle, and  Figure \ref{fig01}(b) the 
climatological mean seasonal cycle, which is the monthly mean value from the total 39 cycles subtracted from the original data. In the former we see that the deviation 
from the climatological mean (red), modulo removal of the seasonal cycle, is much smaller than the mean seasonal change.  This is also reflected in 
Figure \ref{fig01}b, where in the mean seasonal cycle, the temperature changes from approximately $10$ to $27^\circ C$, but the monthly standard deviations are just a few degrees.

\section{Non-autonomous stochastic seasonal climate model}

The intrinsic multiple time scales underlie the mathematical form of a simple climate model in the spirit of that described above as 
\begin{align}
\frac{dx}{dt} = F(x,t)+N(t)\xi+D(\tau),
\end{align}
  where $F(x,t)=F(x,t+T)$, with $T=1$ year, the weather related noise forcing is $N(t)\xi$, with $\xi$ Gaussian white noise, 
  and the long-term forcing is $D(\tau)$, where $\tau=\epsilon t$, with $\epsilon \ll 1$.   Now, let $\bar{x}$ be the climatological mean seasonal cycle and 
  $\bar{D}$ the average of $D(\tau)$, so that 
 \begin{align}
  \frac{d\bar{x}}{dt}=F(\bar{x},t)+\bar{D}, 
 \end{align} 
 where $|F(\bar{x},t)| \gg N(t)$. Hence, the evolution of a fluctuation about the mean, $\eta \equiv x - \bar{x}$, will satisfy
  \begin{align}
  \frac{d\eta}{dt} = a(t)\eta+N(t)\xi+d(\tau),
  \label{eq:main}
 \end{align} 
 where $d(\tau) = D(\tau) - \bar{D}$, and because $|\bar{x}| \gg |\eta|$, we keep only $a(t) = \frac{\partial F}{\partial x}|_{x=\bar{x}}$ where $a(t)=a(t+T)$. Next we ask whether we can construct the time dependent functions $a(t)$, $N(t)$ and $d(\tau)$. 

Typical observations of climate include the mean seasonal cycle $\bar{x}$ and fluctuations $\eta$.  In particular, it is common to develop the seasonal cycle from monthly-averaged data, relative to which weather fluctuations are much shorter and thus described as Gaussian white noise in our stochastic model equation \ref{eq:main}.   Using such data we construct the two periodic functions $a(t)$ and $N(t)$, and the slowly-varying forcing $d(\tau)$\footnote{A detailed derivation is found in the supplementary material of \cite{8}, but we outline it here to make this paper reasonably self-contained.}. On time scales such that the slowly-varying function $d(\tau)$ is approximately constant, equation \ref{eq:main} becomes a periodic non-autonomous equation, $\dot{\eta}=a(t)\eta+N(t)\xi$, which is a generalization of the original Ornstein-Uhlenbeck (OU) process $\dot{\eta} = -\lambda\eta+\sigma\xi$.  Hence, the Floquet exponent $\int_{0}^{T}a(t)dt$ is akin to $-\lambda$ in the OU process, which represents the overall 
stability of a system. However, the key difference is that even with a negative Floquet exponent, which implies overall stability, the actual value of $a(t)$ can be positive
for some sub-periodic time interval.  Hence, temporary instability is a characteristic of the model in equation \ref{eq:main}.  

To understand how we determine $a(t)$ and $N(t)$ from monthly averaged data, we examine how they contribute to the observational statistics. 
The central steps are most easily illustrated in the context of the simple OU process, which has the solution $\eta(t)=\sigma e^{-\lambda t}\int_{0}^{t}e^{\lambda t'}dW'$ where
$dW'$ is a Brownian time increment \cite{18} . In  the long-time limit far from the initial condition, the second moment of $\eta(t)$ is 
\begin{align}
\label{eq:2nd}
 \langle \eta(t)\eta(t+\Delta t)\rangle \simeq \frac{\sigma^2}{2\lambda}e^{-\lambda\Delta t} \equiv \langle \eta^2(t)\rangle e^{-\lambda \Delta t},
\end{align}
where we use $\langle dW'(t') dW''(t'') \rangle = \delta(t'-t'') dt'$ and the definition comes from the fluctuation-dissipation theorem.  
Equation \ref{eq:2nd} leads to 
\begin{align}
 \lambda = -\frac{1}{\Delta t}\text{log}\left[\frac{\langle \eta(t)\eta(t+\Delta t) \rangle}{\langle \eta^2(t) \rangle}\right].
\end{align}
If $\Delta t \ll 1$, we can approximate $\lambda$ as
\begin{align}
\lambda \simeq \frac{1}{\Delta t} \frac{\langle \eta^2(t)\rangle - \langle \eta(t)\eta(t+\Delta t)\rangle}{\langle \eta^2(t) \rangle}, 
\end{align}
and hence stability or instability, $\lambda \lessgtr 0$, depends on $\langle \eta^2(t) \rangle \lessgtr \langle \eta(t)\eta(t+\Delta t)\rangle$. 
This condition is expected to be valid even in the periodic non-autonomous stochastic model equation \ref{eq:main}, {\em mutatis mutandis},
to include the contribution of the long-term forcing, $d(\tau)$, on the auto-correlation $\langle \eta(t)\eta(t+\Delta t)\rangle$.  

Now, if we write equation \ref{eq:main} in a discretized form as 
\begin{align} \label{eq:discrete}
\eta(t+\Delta t)-\eta(t) = a(t)\eta(t)\Delta t + N(t)\Delta W + d(\tau)\Delta t, 
\end{align}
wherein $\Delta t = 1/12$ year, and multiply both sides by 
$\eta(t)$ and take the ensemble average we find
{\color{black}
\begin{align} \label{eq:dis_main}
\langle\eta(t+\Delta t)\eta(t)\rangle - \langle \eta^2(t) \rangle = a(t)\langle \eta^2(t) \rangle\Delta t + d(\tau) \langle\eta(t)\rangle\Delta t.
\end{align}
}

On the other hand, the solution of equation \ref{eq:main} is
\begin{align} \label{eq:sol}
\eta(t) &\simeq \text{exp}\left(\int_{0}^{t}adt'\right)\int_{0}^{t}N(t')\text{exp}\left(-\int_{0}^{t'}ads\right)dW' \nonumber \\
  &+d(\tau)\text{exp}\left(\int_{0}^{t}adt'\right)\int_{0}^{t}\text{exp}\left(-\int_{0}^{t'}ads\right)dt',
\end{align}
where the slowly varying $d(\tau)$ is a constant in the integral.  Using equation \ref{eq:sol} we can prove that $\langle \eta(t)\Delta W\rangle = 0$ and 
$\langle\eta(t)\rangle = d^2(\tau) P(t)$ where $P(t) \equiv \text{exp}\left(\int_{0}^{t}a(t')dt'\right)\int_{0}^{t}\text{exp}\left(-\int_{0}^{t'}a(s)ds\right)dt'$.
We also
write the auto-correlation $\langle \eta(t+s) \eta(t) \rangle$ analytically as
\begin{align} 
 \langle \eta(t+s) \eta(t) \rangle &= \text{exp}\left(\int_{t}^{t+s} a(t')dt'\right)\left(\langle \eta^2(t)\rangle
 - d^2(\tau)P^2(t)\right) \nonumber \\
 &+ d^2(\tau)P(t)P(t+s).
\end{align}
Taking $s = mT$, with $m$ a positive integer and $T$ the period, then this auto-correlation becomes
\begin{align}
\langle \eta(t+mT)\eta(t)\rangle &= e^{-m\lambda}\left(\langle \eta^2(t)\rangle - d^2(\tau)P^2(t)\right) \nonumber \\
&+ d^2(\tau)P^2(t).
\end{align}
For $\lambda$ positive such that $e^{-m\lambda} \ll 1$, for a certain positive integer $m$, then 
\begin{align}
 \langle \eta(t+mT)\eta(t)\rangle \simeq d^2(\tau)P^2(t).
\end{align}
Multiplying both sides of equation \ref{eq:dis_main} by $P(t)$ leads to
\begin{align}
 \left[A(t) - S(t)\right]P(t) = S(t)a(t)P(t)\Delta t + B(t)\Delta t,
 \label{eq:int}
\end{align}
where $A(t) \equiv  \langle \eta(t+\Delta t) \eta(t) \rangle$, $S(t)=\langle \eta^2(t) \rangle$ and $B(t) \equiv  \langle \eta(t+mT)\eta(t)\rangle$.
The definition of  $P(t)$ (following \ref{eq:dis_main}) leads to the relationship $dP(t)/dt=a(t)P(t)+1$, from which we obtain
\begin{align} \label{eq:P}
\frac{dP(t)}{dt} = \frac{1}{\Delta t}\frac{S(t)-A(t)}{S(t)}P(t) + \frac{S(t)-B(t)}{S(t)}.
\end{align}
 {\color{black} (Note that  equation \ref{eq:P} contains the contribution of the long-term forcing $d(\tau)$.)} 
Therefore, with monthly averaged data spanning several decades we can calculate $A(t)$, $S(t)$ and $B(t)$.   {\color{black}Finally, by letting $a(t)P(t)=\frac{dP(t)}{dt}-1$ in equation \ref{eq:int},} $P(t)$ can be calculated from equation \ref{eq:P}, which leads to 
\begin{align}
 a(t) = \frac{1}{P(t)}\left(\frac{dP}{dt} - 1\right).
\end{align}

Now, in order to construct the noise intensity $N(t)$, we multiply both sides of equation \ref{eq:discrete} by $\eta(t+\Delta t)$ 
and take the ensemble average, which  leads to 
\begin{align}
 S(t+\Delta t)-A(t) &= a(t)A(t)\Delta t + N^2(t)\Delta t \nonumber \\
 &+ d^2(\tau)P(t)P(t+\Delta t), 
\end{align}
from which it follows that
\begin{align}
 N^2(t) = \frac{S(t+\Delta t) - A(t)}{\Delta t} - a(t)A(t) - \tilde{B}(t),
\end{align}
where $\tilde{B}(t) = \langle \eta(t+mT+\Delta t)\eta(t)\rangle \simeq d^2(\tau)P(t)P(t+\Delta t)$. 

The procedure is closed by determining the long-time forcing, $d(\tau)$, as a residual viz., 
\begin{align}
d(\tau) = \frac{\int_{t}^{t+T} \frac{1}{N(t')}\left[\frac{d\eta}{dt'}-a(t')\eta(t')\right] dt'}{\int_{t}^{t+T}\frac{1}{N(t')}dt'}
\end{align}

Next we demonstrate the method using monthly averaged climate data. The ensemble average is replaced by time-average. When the data spans several decades, there is sufficient information to calculate monthly statistics. Based on this framework for a single time-series, we extend the method to two different time-series to examine their coupling on seasonal time scales.  
  
\subsection{Application to Tropical Climate Indices}

The eastern tropical Pacific is a key region of upwelling \cite{20}, which is driven by Ekman pumping that forces cold waters to the surface as follows. 
The trade winds (easterly surface winds in the eastern tropical Pacific) exert surface stress on the ocean mixed layer, which drives Ekman pumping. 
When cold waters are brought to the surface, high pressure in the eastern tropical Pacific is enhanced, resulting in an increased east-west pressure difference, further 
intensifying the easterly wind. This positive enhancement is called the ``Bjerknes feedback'' \cite{11}.
The typical state of the Bjerknes feedback in which cold water is brought to the surface in the eastern tropical Pacific, maintains atmospheric 
tropical convective dynamics known as the Walker circulation \cite{21}. However, if the easterly trade winds weaken, the suppression of 
Ekman pumping and the associated upwelling cause warming over the eastern tropical Pacific.  When that warming exceeds a threshold, the 
El-Ni\~{n}o state is operative, dramatically changing the Walker circulation.  Not only does this impact the entire tropical Pacific,  but the influence extends to the 
mid-latitudes and often to the high-latitudes.   When the eastern tropical Pacific is colder than normal and atmospheric pressure is high (low) in the eastern (western) Pacific, the 
La-Ni\~{n}a is operative.  Because both the warm (El-Ni\~{n}o) and cold (La-Ni\~{n}a) states of ENSO have global-scale implications \cite{19}, it is crucial to monitor the state of the eastern tropical Pacific Ocean.  

We take a single time series and construct the coefficients of equation \ref{eq:main} and then demonstrate the reliability by generating the same statistics
as found in the original climate data. The monthly evolution of the tropical climate variability in air-sea interactions is represented by the Nino3.4 and the Indian Ocean Dipole {\color{black}indices} from 1870 to 2018.   In particular, the Nino3.4 index is the Sea Surface Temperature (SST) anomaly within 5S-5N, 120W-170W  \cite{14}. The Nino3.4 index exhibits ENSO through a chaotic oscillation with a period between 2 and 7 years, which is controlled by processes described above. Heuristically similar dynamics are seen in the much smaller tropical Indian Ocean basin, in which the temperature shows an west-east seesaw-like oscillation, called the Indian Ocean Dipole mode (IOD) \cite{16}.  
The IOD is also defined by an index that is the anomalous SST gradient between the western equatorial (50E-70E and 10S-10N) and the
south eastern equatorial (90E-110E and 10S-0N) Indian Ocean \cite{15}. 

\begin{figure}[!h]
\centering\includegraphics[width=3.3in]{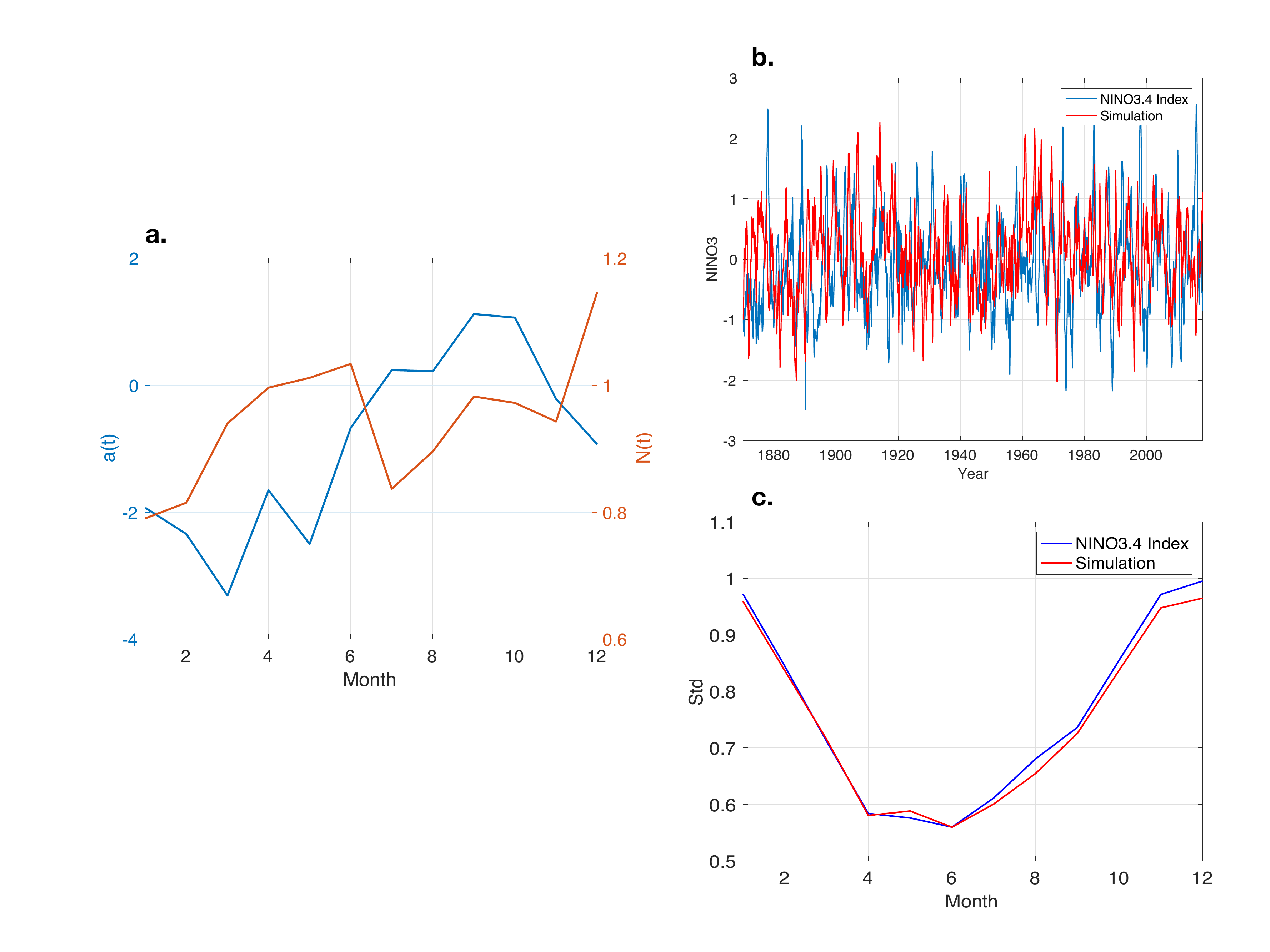}
\caption{Determining the coefficients of the periodic non-autonomous stochastic model, equation \ref{eq:main}, for the El-Ni\~{n}o Southern Oscillation (ENSO). {\color{black} The status of ENSO is monitored by Nino3.4 index defined by Sea Surface Temperature anomaly averaged over 5S-5N, 120W-170W. The monthly-averaged Nino3.4 index data used in our analysis 
spans 149 years from 1870 to the present, and thus we have an ensemble of 149 samples to compute second-order statistical moments for each month. }
(a) The values of $a(t)$ and $N(t)$. (b) The comparison between the time-series of the Nino 3.4 index (blue line) and a stochastic realization of the stochastic model equation \ref{eq:main} (red line) without the long term forcing, $d(\tau)$. (c) is the monthly standard deviation comparison. 
}
\label{fig02}
\end{figure}

In figure \ref{fig02}a we show $a(t)$ (blue line) and $N(t)$ (orange line) for the Nino3.4 index and in figure \ref{fig02}b we compare the
original time-series of the Nino3.4 index from 1870 to 2018 (blue line) with a stochastic realization of the model (red line). 
Here for the model simulation we exclude the long-term forcing, $f(\tau)$, which does not contribute to the seasonal statistics. 
Figures \ref{fig02}c,d show the monthly standard deviations and power spectrum respectively.  These comparisons show that the model regenerates the monthly statistics of the original data.

The seasonal variability of ENSO can be discussed in terms of the interaction between the two periodic functions $a(t)$ and $N(t)$. 
The monthly stability of ENSO is described by the sign of $a(t)$; when it is positive (negative), the response of the system is to increase (decrease). 
{\color{black} The effect of instantaneous noise, $N(t)\xi$, can be enhanceed or suppressed by the monthly stability}.  
The monthly stability $a(t)$ is positive (negative) from July to November (from December to June).  
Hence the variability of the response, $\eta$, is expected to be maximal in November when $a(t)$ changes sign from positive to negative.  
However, the noise intensity sharply increases during December, which shifts the timing of the maximum from November to December. 
The warm (El-Ni\~{n}o) and cold (La-Ni\~{n}a) states of ENSO are statistical outliers, which are more probable near the end of a year 
when $a(t)$ changes sign.  Indeed, the origin of the names El-Ni\~{n}o and La-Ni\~{n}a, is associated with their occurrence near Christmas.  
This ``phase locking'' is interpreted within the framework of our stochastic model equation \ref{eq:main}, where the seasonal
the Bjerknes feedback, $a(t)$, interacts with the high frequency noise, $N(t)$, to generate the maximum ENSO variability near the end of year.

\begin{figure}[!h]
\centering\includegraphics[width=3.3in]{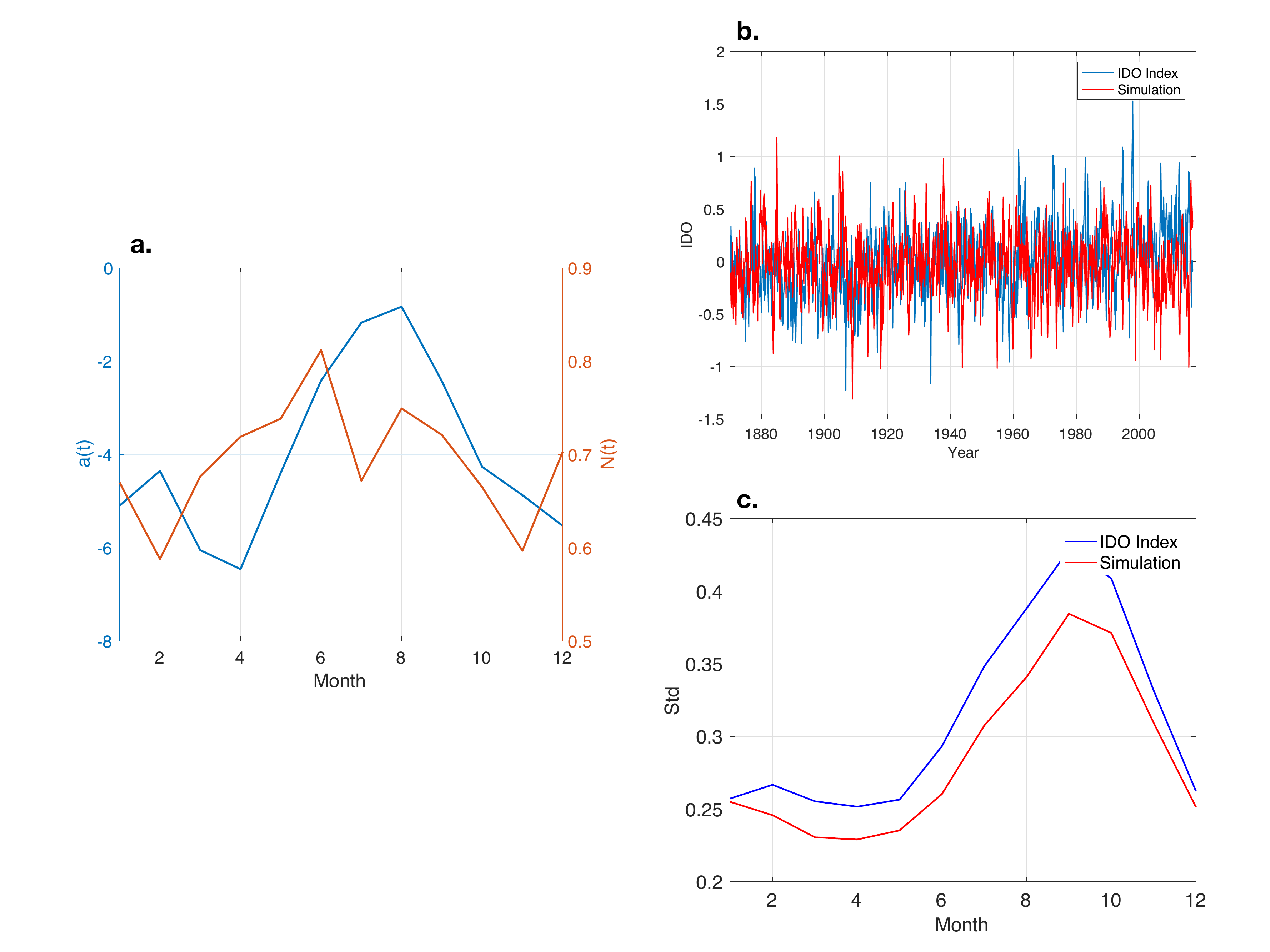}
\caption{ {\color{black}In the Indian Ocean, there is an air-sea interaction pattern similar to that of ENSO, which is called the Indian Ocean Dipole (IOD).  (It is 
also referred to as the Indian Ni\~{n}o). In brief, the IOD is the irregular oscillation of sea surface temperature difference between the western and the eastern Indian Ocean, which is indexed as the anomalous SST gradient between the western equatorial (50E-70E and 10S-10N) and the
south eastern equatorial (90E-110E and 10S-0N) Indian Ocean. The coefficients of the periodic non-autonomous stochastic model for the IOD are constructed and 
shown in the same manner is in figure \ref{fig02}.} }
\label{fig03}
\end{figure}

In figure \ref{fig03} we apply the same logic to the IOD.  However, unlike ENSO, the IOD does not exhibit unstable deterministic forcing during the seasonal cycle, but has instead a strong minimum during summer (blue line in \ref{fig03}(a)). Thus, the combination of this weakened stability and the noise leads to maximum variability in September. 
The overall variability of the IOD is smaller than that of ENSO, which suggests that the Bjerknes feedback in the former is weaker than that in the latter. Nonetheless, the stochastic model reliably reproduces the main features and the monthly statistical dynamics of both.

The extent to which the lower dimensional stochastic model of equation \ref{eq:main} does not reproduce the behavior of these two climate {\color{black}phenomena} may be associated with their dynamical coupling through the Walker circulation. Therefore, to represent the interaction of {\color{black}these two phenomena}, we extend the approach to two-dimensions using coupling functions.    

\section{Coupling Functions in two-dimensions}
\label{sec:coupling}

We consider a simple linear coupling approach introduced by \cite{22,23}, which in our framework is
\begin{align}
 &\frac{d\eta_1}{dt} = a_1(t)\eta_1+N_1(t)\xi_1(t)+b_{12}(t)\left(\eta_2-\eta_1\right) \qquad \text{and} \nonumber \\
 &\frac{d\eta_2}{dt} = a_2(t)\eta_2+N_2(t)\xi_2(t)+b_{21}(t)\left(\eta_1-\eta_2\right),
 \label{eq:coupling}
\end{align}
where $a_{1,2}(t)$ and $N_{1,2}(t)$ are the monthly stability and the noise intensity of the variabilities $\eta_{1,2}$, and 
$b_{12,21}(t)$ represent the coupling influence from other {\color{black}phenomenon}. Following the method described above 
for the 1-D model, we construct the periodic functions $a_{1,2}(t)$, $N_{1,2}(t)$ and $b_{12,21}(t)$
using monthly variance and covariance data. 

First we multiply both sides of the time-evolution equation for $\eta_1(t)$ in equation \ref{eq:coupling} by $\eta_{1}(t)$ and $\eta_2(t)$ respectively, and then take the 
ensemble average to give
\begin{widetext}
\begin{align}
&\langle \eta_1(t) \frac{d\eta_1}{dt}\rangle = a_1(t)\langle \eta^2_1(t)\rangle + b_{12}(t)\left[\langle \eta_1(t)\eta_2(t)\rangle - \langle \eta^2_1(t)\rangle\right] \qquad \text{and}\nonumber \\
&\langle \eta_2(t) \frac{d\eta_1}{dt}\rangle = a_1(t)\langle \eta_1(t)\eta_2(t)\rangle + b_{12}(t)\left[\langle \eta^2_2(t) \rangle - \langle \eta_1(t)\eta_2(t) \rangle\right],
\end{align}
\end{widetext}
where
\begin{align}
 \langle \eta_{1,2}(t) \frac{d\eta_1}{dt} \rangle \simeq \frac{1}{\Delta t} \left(\langle \eta_{1,2}(t) \eta_1(t+\Delta t)\rangle -\langle \eta_{1,2}(t)\eta_1(t)\rangle\right).
 \end{align}
Here again $\Delta t = 1/12$ yr for monthly averaged climate data.  
The above equations are represented by $\boldsymbol{A_1(t)}X_1(t) = Q_1(t)$, in which 
\begin{align}
\boldsymbol{A_1(t)} & = 
  \begin{bmatrix}
     \langle \eta^2_1(t) \rangle & \langle \eta_1(t)\eta_2(t) \rangle - \langle \eta^2_1(t)\rangle \\
     \langle \eta_1(t)\eta_2(t) \rangle & \langle \eta^2_2(t) \rangle - \langle \eta_1(t)\eta_2(t) \rangle 
  \end{bmatrix}, 
\end{align}
where $X_1(t) = [a_1(t) \;\; b_{12}(t)]^T$ and $Q_1(t) = [\langle \eta_1(t) \frac{d\eta_1}{dt}\rangle \;\;  \langle \eta_2(t) \frac{d\eta_1}{dt}\rangle]^T$.
It is possible to construct $\boldsymbol{A_1(t)}$ and $Q_1(t)$ for each month from the data and thereby
obtain $X_1(t) = \boldsymbol{A^{-1}_1(t)} Q_1(t)$.  Similarly, we obtain $X_2(t) = \boldsymbol{A^{-1}_2(t)} Q_2(t)$, in which 
\begin{align}
\boldsymbol{A_2(t)} & = 
  \begin{bmatrix}
     \langle \eta^2_2(t) \rangle & \langle \eta_1(t)\eta_2(t) \rangle - \langle \eta^2_2(t)\rangle \\
     \langle \eta_1(t)\eta_2(t) \rangle & \langle \eta^2_1(t) \rangle - \langle \eta_1(t)\eta_2(t) \rangle 
  \end{bmatrix}, 
\end{align}
where $X_2(t) = [a_2(t) \;\; b_{21}(t)]^T$ and $Q_2(t) = [\langle \eta_1(t) \frac{d\eta_2}{dt}\rangle \;\;  \langle \eta_2(t) \frac{d\eta_1}{dt}\rangle]^T$.

To determine the noise intensities $N_1(t)$ and $N_2(t)$, we multiply both sides of equation \ref{eq:coupling} by $\eta_{1,2}(t+\Delta t)$, which leads to
\begin{widetext}
\begin{align}
N^2_{1,2}(t) = \langle \eta_{1,2}(t+\Delta t)\frac{d\eta_{1,2}}{dt}\rangle - a_{1,2}(t)\langle \eta_{1,2}(t)\eta_{1,2}(t+\Delta t)\rangle 
   -b_{12,21}(t)\left(\langle \eta_{1,2}(t+\Delta t)\eta_{2,1}(t)\rangle -\langle \eta_{1,2}(t+\Delta t)\eta_{1,2}(t)\rangle\right)
\end{align}
\end{widetext}

The six periodic functions $a_{1,2}(t)$, $N_{1,2}(t)$ and $b_{12, 21}(t)$ can be calculated from the second-order statistics in the observed variables.
Physically, the $a_{1,2}(t)$ represent the {\em local stability} of a system, whereas the overall stability is given by $a_{1,2}(t)-b_{12,21}(t)$, showing that the coupling
influences the global stability. At the same time, the coupling functions $b_{12,21}(t)$ provide a forcing $b_{12,21}(t)\eta_{2,1}$
to the other variable $\eta_{1,2}$. The high frequency noise forcing $N_{1,2}(t)\xi_{1,2}(t)$ captures effects that are not represented by the interactions.

\subsection{Dynamic interaction between ENSO and IOD}

We now apply the approach to the Nino3.4 and the IOD indices to determine $a_{1,2}(t)$, $N_{1,2}(t)$ and the coupling functions $b_{12,21}(t)$. 
The interaction of these {\color{black}phenomena} cannot be entirely revealed by the monthly covariance, 
which does not give information about the {\em direction} of influence between the two phenomena. For example, here, $b_{12}(t)$ is the influence of IOD upon ENSO.   

\begin{figure}[!h]
\centering\includegraphics[width=3.4in]{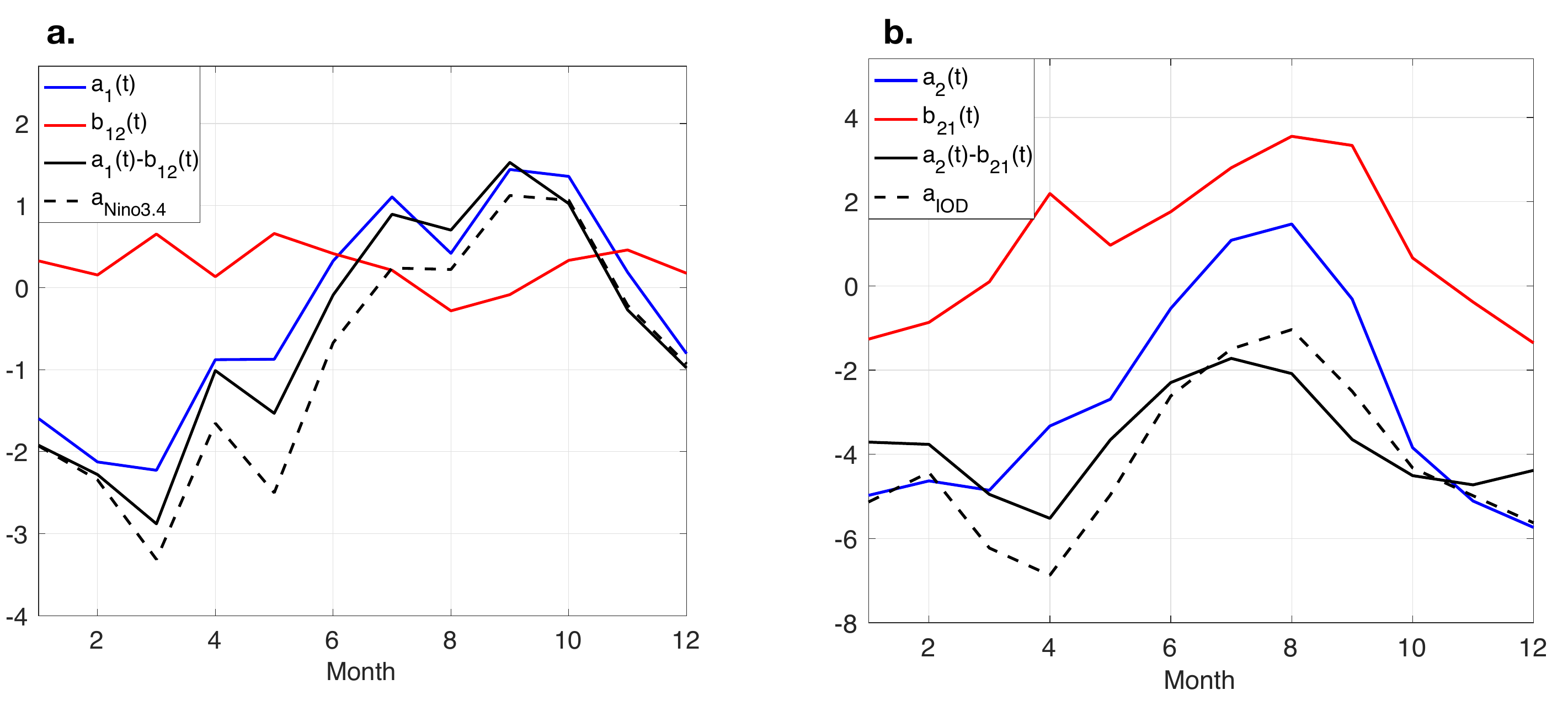}
\caption{ {\color{black}Using the Nino3.4 and IOD indices, the coupling equation, Eq. \ref{eq:coupling}, is applied to ENSO and the IOD to quantify their seasonal interaction. Seasonal stabilities, $a_{1,2}(t)$, coupling intensities, $b_{12, 21}(t)$,
and the intensity of the noises, $N_{1,2}(t)$, are calculated following the procedures in section \ref{sec:coupling}.}
(a) The seasonal stability for ENSO, $a_1(t)$ (blue), 
the influence of IOD on ENSO, $b_{12}(t)$ (red), $a_1(t)-b_{12}(t)$ (thick black) and $a_\text{Nino3.4}$ (Dashed black). (b) The same as (a) but for $a_2(t)$, the influence of ENSO on IOD, $b_{21}(t)$, $a_2(t)-b_{21}(t)$ and $a_\text{IOD}$.
}
\label{fig04}
\end{figure}

We show in figure \ref{fig04}a the results for $a_1(t)$ (blue), $b_{12}(t)$ (red) and the comparison between
$a_1(t)-b_{12}(t)$ (black) and $a_\text{Nino3.4}$ (dashed black), with the stability based only on the Nino3.4 index data.  We find that the stability $a_1(t)-b_{12}(t)$ in the two dimensional system is similar to that in the one-dimensional system, $a_\text{Nino3.4}$.  During the seasonal cycle, $b_{12}(t)$ straddles the origin, whereas $a_1(t)$ varies from -2 to 1.5, suggesting that the influence of the IOD upon ENSO does not control the seasonal variability of the latter.  
Only in August does the coupling $b_{12}$ become slightly negative, so that the IOD provides a weak positive forcing to ENSO and enhances its stability. However, the overall impact of the IOD upon the ENSO is not significant. This is consistent with modeling studies that show only extreme IOD events impact ENSO \cite{24}. 

We show in figure \ref{fig04}b the results for $a_2(t)$ (blue), $b_{21}(t)$ (red) and the comparison between
$a_2(t)-b_{21}(t)$ (black) and $a_\text{IOD}$ (dashed black). Here, unlike the influence of the IOD on ENSO in figure \ref{fig04}a, the influence of ENSO
upon the IOD, is non-negligible. Indeed, the coupling function $b_{21}(t)$ (red) has large amplitude seasonal variations with a large positive value in August. 
The similar magnitudes of $a_2(t)-b_{21}(t)$ and $a_\text{IOD}$ suggests how the coupling $b_{21}(t)$ controls the overall stability, $a_\text{IOD}$.
Clearly, ENSO stabilizes the IOD while simultaneously providing positive forcing to the time-evolution 
of the IOD, implying the strong influence of the ENSO upon the IOD \cite{25}.

\begin{figure}[!h]
\centering\includegraphics[width=3.4in]{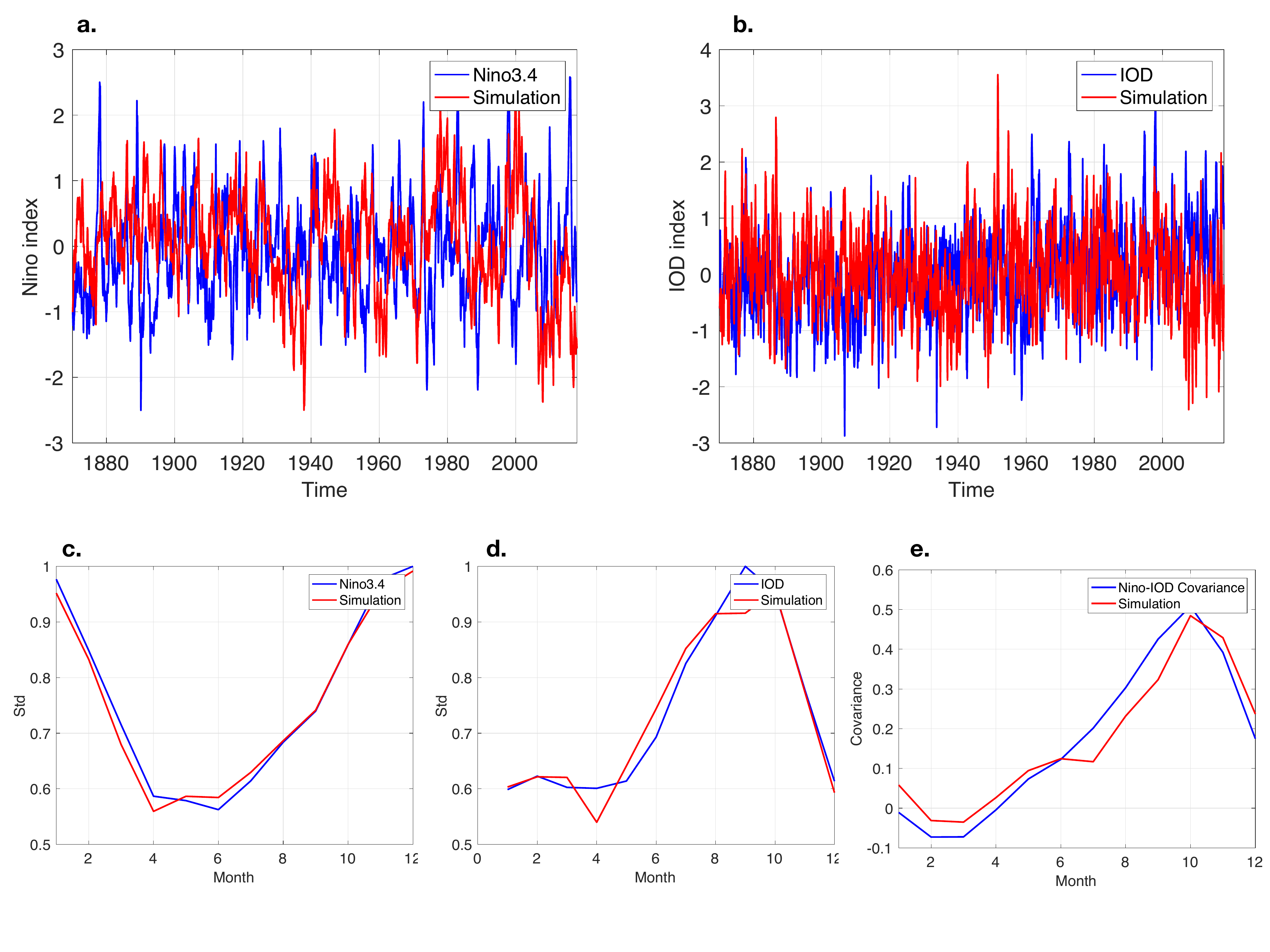}
\caption{The two-dimensional stochastic model (Eq. \ref{eq:coupling}) is simulated and compared with data. In (a) and (b) the simulated $\eta_1$ and $\eta_2$ (red lines)
are compared to the original Nino3.4 and IOD indices (blue lines). The monthly standard deviations 
of (c) $\eta_1$ and the Nino3.4 index, and (d) $\eta_2$ and the IOD index. (e) The covariance 
between $\eta_1$ and $\eta_2$ is compared to that between the Nino3.4 and the IOD indices.}
\label{fig05}
\end{figure}

Finally, having determined the coefficients, we simulate the 2-dimensional model to compare the results with the data.  Figure \ref{fig05} shows the time-series of ENSO (a) and the IOD (b), monthly standard deviations (c, d) and covariances (e).  The statistical structure is well captured by the stochastic simulation. 
The standard deviations calculated from ensemble simulations for ENSO and the IOD (red) are compared with those from the data (blue). Whilst there are systematic deviations, the overall seasonal evolution of the standard deviations is well captured, as are the covariances.  This is particularly important in that the maximal covariance is in October, when the seasonal variability of the IOD is also maximal, suggesting a strong influence of ENSO on the IOD. 

\vspace*{-5pt}

\section{Conclusion}

The climate system can be viewed in terms of the interaction of many subsystems on multiple time and length scales.  Some subsystems can be approximately treated as being spatially localized on some time scales, but they evolve in consequence of forcing on multiple other time scales.  Other subsystems are less well treated as being spatially localized on the same time scales and are therefore coupled to some degree with others.  
{\color{black} We have viewed these as limits of a generalized multiple-time-scale non-autonomous stochastic treatment and proposed a model of their coupling.  We provide a method to (a) derive analytical expressions for the coefficients of the one- and two-dimensional stochastic differential equations, and (b) determine their values from the climate data.  We applied the method to the Nino3.4 and IOD indices, which describe the El-Ni\~{n}o Southern Oscillation and Indian Ocean Dipole. The approach successfully allows us to infer important causal linkages.}

This approach allows one the ability to simulate the dynamics of isolated and coupled climate subsystems.  For example, by constructing coupling functions, we have demonstrated the dominant influence of ENSO on the IOD during the mature period--the summer--of the IOD. Moreover, we show that the IOD has little influence on ENSO excepting in the case of an extreme IOD event. This simple stochastic dynamical picture yields these, and other results, that are consistent with the approaches of global climate model simulations and climate reanalysis
data. 

The simple coupling function approach introduced here demonstrates the utility of replacing the practice of using the covariance itself to detect climate subsystem interactions.  Namely, this approach provides a stochastic model that contains the the second-order statistics seen in data. In particular, the coupling function method could be a main research tool in studying climate teleconnections.   The examples provided here might act as a useful starting point, with next steps being generalization to a multi-dimensional stochastic model to represent the climate as a high dimensional network.   


\vspace*{-7pt}

\vskip2pc


\end{document}